\documentclass[letterpaper, 10 pt, journal, twoside]{IEEEtran}
\usepackage{cite}
\usepackage[pdftex]{graphicx}
\usepackage{amsmath}
\usepackage{url}
\usepackage{commath}
\usepackage{gensymb}
\usepackage{doi}
\newcommand{\R}{\mathcal{R}}

\IEEEoverridecommandlockouts                              

\title{Can the COVID-19 epidemic be controlled\\ on the basis of daily test reports?}

\author{Francesco Casella$^{1}$
\thanks{$^{1}$Francesco Casella is with Dipartimento di Elettronica, Informazione e Bioingegneria, Politecnico di Milano, Piazza Leonardo da Vinci 32, 20133 Milano, Italy; tel. +39 02 2399 3465
        {\tt\small francesco.casella@polimi.it}}%
}

\begin{document}

\maketitle

\thispagestyle{empty} 

\begin{abstract}
This paper studies if and to which extent COVID-19 epidemics can be controlled by authorities taking decisions on public health measures on the basis of daily reports of swab test results, active cases and total cases. A suitably simplified process model is derived to support the controllability analysis, highlighting the presence of very significant time delay; the model is validated with data from several outbreaks. The analysis shows that suppression strategies can be effective if strong enough and enacted early on. It also shows how mitigation strategies can fail because of the combination of delay, unstable dynamics, and uncertainty in the feedback loop; approximate conditions based on the theory of limitation of linear control are given for feedback control to be feasible.
\end{abstract}


\section{Introduction} \label{sec:Introduction}
\IEEEPARstart{T}{he} first outbreak of the COVID-19 \cite{JAMA} virus epidemic took place in China, starting at the end of 2019, and has since then caused a global pandemic with disruptive effects on public health, social life, and the economy. The pandemic will likely spark a large number of studies to understand its behaviour and to determine effective control strategies.

A wide range of mathematical models have been proposed to describe the dynamic evolution of epidemics, starting from the seminal paper \cite{KermackMcKendrick1927}, and including a wide range of possibly quite sophisticated models, see e.g. \cite{Hethcote2000} for a comprehensive review. The analysis of these models allows to predict the evolution of the disease over time, its asymptotic behaviour (e.g. endemic disease equilibria vs. eradication), and most importantly how it depends on the model parameters. 

Epidemiological models are widely used to design vaccination and treatment strategies based on optimal control, see, e.g. \cite{SharomiMalik2017} and references therein. They can also be used to design feedback vaccination strategies \cite{DeLaSenAlonsoQuesada2011}, or even feedback strategies combining different actions such as vaccination, treatment, and culling \cite{DeLaSenEtAl2017}. Some studies take into the account the feedback effects of behavioural changes in the evolution of an epidemic, see \cite{PolettiEtAl2012} and reference therein.

Most models employed to study control strategies are formulated in terms of ordinary differential equations, e.g. the classical SIR and SEIR models and their variants. In some cases, time delays are also included in the model, to account for the incubation time, see, e.g., \cite{YoshidaHara2007}, \cite{ZamanEtAl2009}.

Detailed models of the COVID-19 outbreak have started to appear in the literature. With reference to the outbreak in Italy, \cite{GiordanoEtAl2020} proposes an extension of the classic SIR model with eight state variables, while \cite{GattoEtAl2020} presents a spatially resolved model with nine state variables for each of the 107 provinces of the country. Both models confirmed the appropriateness of the public measures taken by the Italian authorities to contain the virus outbreak. A highly detailed epidemiological model of the UK was used in \cite{Ferguson2020a} to predict possible outcomes of the virus outbreak and to suggest the adoption of a suppression policy. 

The report \cite{Flaxman2020} attempts to estimate the effects of non-pharmaceutical interventions (NPIs, i.e., public health measures) onto the relative reduction of the reproduction number $\R_t$ of COVID-19, by applying Bayesian methods to data from 11 European countries. Given the estimates of the initial reproduction number $R_0$, a reduction by at least 60-70\% or more is necessary to suppress exponential growth. The main result of \cite{Flaxman2020} is that lockdown leads to an average reduction of $\R_t$ by 50\%, school closure by 20\%, other measures around 10\%. However, 95\% confidence intervals on the reduction factors are huge, e.g. 10\% to 80\% reduction for lockdown, 0\% to 45\% reduction for school closure, severely undermining their predictive power. This problem is inherent to the requirement of a large enough data set to be statistically significant, which requires to put countries with very different social habits and very different interpretations of the same measure (e.g. lockdown) in the same data set.

The use of feedback control theory has been advocated early on as a powerful tool to support the management of the COVID-19 outbreak \cite{StewartEtAl2020}. Unfortunately, most of the existing literature on the control of epidemics involves vaccines or treatments, which are currently not available for COVID-19. Some innovative feedback control strategies have been proposed in preprints at the time of this writing, e.g. \cite{BinEtAl2020}, which proposes a feedback mitigation strategy based on fast lockdown cycles controlled by a supervisory loop, or \cite{MuellerEtAl2020}, advocating a strategy based on massive random testing.

The aim of this paper is to assess the controllability of the COVID-19 outbreak, assuming that the population is sufficiently well mixed and that the decisions of public health measures by the authorities are based on daily reports of positive swab tests, active cases, and total cases. To this aim, a suitably simplified model is presented, which is specifically aimed at capturing the fundamental dynamics of the process which is relevant for feedback control, which turns out to be heavily affected by time delay.

The main result of the analysis is twofold. On one hand, \emph{suppression}  strategies can be effective if enacted early on and with strong enough measures. On the other hand, \emph{mitigation} strategies  turn out to be infeasible if the reproduction number is significantly higher than one, and are in any case limited by the time delays in the feedback loop.

The paper is structured as follows. In Section \ref{Sec:Modelling}, a control-oriented model of the epidemic is introduced and validated against data from the outbreaks in different countries. In Section \ref{Sec:Control}, the two above-mentioned strategies are analysed in terms of feedback control. Section \ref{sec:Conclusions} draws conclusions from the control-theoretical analysis with some recommendations for decision makers and future research.

\section{Modelling}
\label{Sec:Modelling}
\subsection{Derivation of the model}
Models of the COVID-19 epidemic such as those mentioned in Sect. \ref{sec:Introduction} are based on first principles, in the sense that their equations describe the time evolution of different categories of subjects, based on the known mechanisms of infection, recovery, and care of patients. However, their behaviour is ultimately decided by the values of coefficients that must be identified from experimental data, which is preciously scarce in the case of a new disease such as COVID-19. The quality and homogeneity of data used to tune those models are also often  questionable: different countries adopt different standards for swab testing, possibly changing them over time; some data get lost because of clerical errors; some countries or regions may report lower numbers than real because of political pressures. Even bona-fide reports may fail to provide reliable data, as revealed by the mismatch between official COVID-19 deaths and additional numbers of deaths on municipal records in previous years. The actual effects of NPIs are still quite uncertain, see \cite{Flaxman2020}.

Public policies based on such models cannot thus be applied blindly, but must be adapted and corrected based on the observed outcome. Indicators used by public decision makers include daily reports of a) new positive swab tests, b) current number of infected subjects, and c) total number of reported cases. The crucial question is then: \emph{is feedback control feasible at all in such a system}?

In order to answer this question, a suitably simplified model of the epidemic is derived here to capture the fundamental dynamics that is relevant for the design of the feedback policy, in particular the dynamic relationship between NPIs and the response of the three above-mentioned indicators. The starting point is the basic SEIR model \cite{Hethcote2000} with the addition of a further compartment $L$:
\begin{align}
& \frac{dS}{dt} = - \frac{\beta I S}{N} \label{eq:S}\\
& \frac{dE}{dt} =  \frac{\beta I S}{N} - \epsilon E \label{eq:E} \\
& \frac{dI}{dt} = \epsilon E - \gamma {I} \label{eq:I}\\
& \frac{dL}{dt} = \gamma I - \delta L \label{eq:L} \\
& \frac{dR}{dt} = \delta L \label{eq:R}
\end{align}
were $N$ is the total population, $S$ is the number of Susceptible individuals, $E$ is the number of Exposed individuals, that have caught the infection but are not yet infectious, $I$ is the number of Infectious individuals, $L$ is the number of subjects which are still iLl, but are no longer infectious due to hospitalization, quarantine, or just because infected subjects are mostly infectious during the first few days after the end of the latency period \cite{HeEtAl2020}, and $R$ the number of recovered resistant subjects.

The parameter $\beta$ accounts for the likelihood of infection per unit time; $\epsilon$ is the inverse of the average latency time before one becomes infectious, $\gamma$ is the inverse of the average time infectious subjects spend by actually infecting other people, and $\delta$ is the inverse of the average time subjects remain ill but without infecting others. Given the short time spans involved and the relatively low mortality rate, deaths and births can be neglected, as well as immigration and emigration, that are restricted during the outbreak.

The features of the COVID-19 virus, coupled with the unavailability of effective treatments at the time of this writing, are such that allowing more than a few percent of the population to be infected at any point in time is unacceptable, as doing so would lead to a collapse of the public health system, particularly with reference to the significant fraction of infected subjects needing intensive care to survive the acute respiratory syndrome that the virus can cause. This fact, coupled with the fairly long recovery time (about one month), means that even the worst outbreaks in Western Europe are currently estimated to have infected less than 10\% of the population after a few months in the course of the epidemic. This allows to consider $S(t)/N \approx 1$ and get rid of Eq. \eqref{eq:S}. In fact, a significant portion of the population may be not susceptible \emph{a priori}, e.g. due to genetic reasons. However, absent any concrete evidence of this fact, the precautionary principle suggest to consider the worst case $S(t)/N = 1$.

Assuming then a constant value of $\beta$, the three eigenvalues of system \eqref{eq:E}-\eqref{eq:L} are $-\delta$ and the two roots $p$ and $r$ of 
\begin{equation}
s^2 + (\epsilon + \gamma)s -\epsilon (\beta - \gamma) = (s-p)(s-r).
\end{equation}
If $\beta > \gamma$, there is one negative eigenvalue $p$ and one positive eigenvalue $r$. Assuming that the negative exponential mode has already died out, the solution of \eqref{eq:E}-\eqref{eq:I} is then:
\begin{align}
& I(t) \approx I(0) e^{rt},\label{eq:exp1} \\
& E(t) \approx \frac{\beta I(0)}{r+\epsilon} e^{rt} \label{eq:exp2}
\end{align}
The doubling time of $I(t)$ is $T_d = log(2)/r$. Under the assumption that $S \approx N$, we can approximate the current reproduction number $R_t = \beta/\gamma$. This formula is not exact during transients, when the number of infectious subjects changes over time, but allows to later obtain some interesting synthetic results, though with some approximation.

In contexts where massive testing was carried out, it was found that about 40\% of positive tested subjects are entirely asymptomatic, despite being infectious \cite{OranTopol2020}, \cite{LavezzoEtAl2020}, which makes COVID-19 particularly insidious. This suggests that a similar fraction of the infectious subjects goes unnoticed and escapes the testing process in the general population. Furthermore, swab tests are affected by more than 20\% false negatives \cite{KucirkaEtAl2020}, and some mildly symptomatic subjects may also end up not being tested. The ratio $\alpha$ between the number the infectious subjects $I(t)$ at a certain time $t$ and the number of infectious subjects $I_t(t)$ at the same time $t$ that will eventually get tested positive is then likely between two and three. On the other hand, $\alpha$ is only relevant to determine when the ratio $S(t)/N$ starts decreasing significantly below one, providing some degree of herd immunity, as all other important indicators, namely the mortality ratio, the ratio of hospitalized subjects and the ratio of subjects requiring intensive care, are all referred to $I_t(t)$.

Assuming that $\alpha$ is constant, one can use Eqs. \eqref{eq:E}-\eqref{eq:R} to also describe the dynamics of the fraction of subjects that are eventually tested positive through the various stages of the disease, $E_t(t)$, $I_t(t)$, and $L_t(t)$.

In most cases, subjects are only tested after they show serious symptoms, which happens on average $\tau_t$ days after they have become infectious. The lab processing also introduces a delay $\tau_r$ before reports are available. Although in principle it is possible to provide the results of the test in a few hours, the average reporting time is usually much longer because of limited equipment availability, up to one week or more.

The NPIs mentioned earlier (lockdown, school closures, etc.) reduce the rate of infection $\beta$, hence the current reproduction number $\R_t \approx \beta / \gamma$. These measures are varied and can be applied progressively. 
We can then assume that the time-varying parameter $\beta$ is in fact a function of a representative manipulated variable $u(t)$, with $u$ indicating the intensity of adopted public health measures on a scale from 0 (no intervention) to 1 (full lockdown and isolation of all individuals). The $\beta(\cdot)$ function is thus monotonously decreasing from the value $\beta_0$, when no social restrictions are enforced, to zero, corresponding to the total isolation of each person in the contry. Of course $\beta$ is also a function of other unknown factors $d(t)$ that act as disturbances on the system, e.g. mutations of the virus or changes in social behaviour which are not directly mandated by the authorities. Considerable uncertainty is involved in the estimation of the effects of different interventions in terms of reduction of $\beta$ or, equivalently, of $\R_t$, see \cite{Flaxman2020}, hence $\beta(\cdot)$ is also uncertain.

The control-oriented model can thus be formulated as a state-space system with output delays:
\begin{align}
\frac{dE_t(t)}{dt} &= \beta(u(t),d(t)) I_t(t) - \epsilon E_t(t) \label{eq:Etf} \\
\frac{dI_t(t)}{dt} &= \epsilon E_t(t) - \gamma  I_t(t) \label{eq:Itf} \\
\frac{dL_t(t)}{dt} &= \gamma I_t(t) - \delta L_t(t) \label{eq:Ltf} \\
\frac{dT_t(t)}{dt} &= \epsilon E_t(t) \label{eq:Ttf} \\
N_r(t) &= \epsilon E_t(t - \tau_m) \label{eq:Nrf}\\
A_r(t) &= I_t(t - \tau_m) + L_t(t - \tau_m) \label{eq:Arf} \\
T_r(t) &= T_t(t - \tau_m) \label{eq:Trf},
\end{align}
where $\beta(u,d)$ is an uncertain function, $\epsilon$, $\gamma$, $\delta$ are uncertain constant parameters, $\tau_t$, $\tau_r$ are uncertain parameters, $\tau_m = \tau_t + \tau_r$ is the overall measurement delay, and $T_t(t)$ computes the cumulative number of positive tested subjects.

The first measured variable of the process is the number of new daily reported cases $N_r(t)$, which is affected by the overall delay $\tau_m$. The second measured variable is the number of reported active cases $A_r(t)$, i.e. the number of subjects for which a positive test report has been received and a double negative test has not yet been issued to certify their recovery. As $\tau_t$ and $\tau_r$ are similar (several days), $2\tau_r \approx \tau_t + \tau_r = \tau_m$, leading to $A_r(t) = I_t(t-\tau_m) + L_t(t -\tau_m)$. The third measured variable is the total cumulated number of reported positive swab test reports $T_r(t) = T_t(t - \tau_m)$. 

Note that the model \eqref{eq:Itf}-\eqref{eq:Trf} has a time delay on the output equations. Since input/output dynamics only will be considered in the next Section, an equivalent representation could be used where the delay is applied to the input instead. 
\subsection{Validation and Tuning}
\label{sec:Validation}
The goal of the model is to describe the dynamic response of the $N_r(t)$, $A_r(t)$, and $T_r(t)$ indicators to the application of NPIs by central authorities, described by changes in $u(t)$. Four outbreaks cases were selected, all characterized by step changes of $u(t)$ at the central government level, in order to make the validation easier: China, with data taken from \cite{JAMA}, Italy, France, and the UK, with data taken from \cite{Worldometer}, which reports data from national authorities.

In the case of Italy and UK, some partial restrictions were introduced first, causing a noticeable delayed reduction of the exponential increase rate of new cases, then a full lockdown was prescribed, whose effect was to change the positive exponential growth of new cases into a negative exponential decay after some delay. We then assume that $\beta = \beta_0$ at $t = 0$, then $\beta$ undergoes a step reduction to $\beta = \rho_1 \beta_0$ at $t = t_1$ and a further step reduction to $\beta = \rho_2 \beta_0$ at time $t = t_2$. In the case of China and France no NPIs were enforced before full lockdown, yet a reduction of the exponential growth rate well ahead of the effects of lockdown is clearly visible, possibly due to social feedback effects or other disturbances. Two step reduction were thus applied also in those cases.

The parameters of the model were tuned manually to obtain a good fit with the available data. In particular, $\tau_m$ is easily tuned to match the delay between the lockdown and the peak of $N_r(t)$, $\R_t = \beta/\gamma$ and $\rho_1$ determine the growth rates in the pre-lockdown behaviour of $N_r(t)$, $\rho_2$, $\gamma$ and $\epsilon$ determine the shape and decay rate of the post-lockdown behaviour of $N_r(t)$. $T_r(t)$ is the integral of $N_r(t)$, so fitting it helps refining the parameter tuning considering the noisy nature of $N_r(t)$, which is also affected by a weekly oscillation due to repeating lab schedules. Finally, $\delta$ is tuned to match the peak of active cases $A_r(t)$, which is much wider and delayed than the peak of $I_r(t)$.  A detailed error analysis was not performed and could be the subject of future work considering a more extensive data set; in any case, high parameter accuracy is not required to support the forthcoming analysis.

The values obtained for the four outbreaks are reported in Table \ref{tab:parameters}. They are fairly consistent with each other and are compatible with the ones reported in \cite{GattoEtAl2020} and \cite{Worldometer}. The initial reproduction number $\R_0$ and the current reproduction numbers $\R_1$ and $\R_2$ computed after each change of $\beta$ are reported, as well as the doubling times $T_{d0}$ and $T_{d1}$ of the unstable mode before and after the first change of $\beta$, and the maximum reproduction number $\R_l$ corresponding to the controllability limit derived in Section \ref{sec:control_feasibility}.  

\begin{table*}[t]
\begin{center}
\caption{Model parameters (time constants in days)}
\label{tab:parameters}
\begin{tabular}{c|c|c|c|c|c|c|c|c|c|c|c|c|c}
\quad  & Period &  $\beta_0$ & $1/\gamma$ & $1/\epsilon$ & $1/\delta$ & $ t_1, \rho_1$ & $ t_2, \rho_2$ & $\R_0$ & $\R_1$ & $\R_2$ & $T_{d0}$  &  $T_{d1}$ & $\tau_m$ \\
\hline
China  & 18/01/2020 -- 11/02/2020 & 1.6     & 2.5     & 5.0       & N/A     & 0,  0.63  & 5,  0.160      &  4.0 & 2.5  & 0.64 & 2.5   &  4.3   & 12 \\
Italy  & 22/02/2020 -- 01/05/2020 & 1.3        & 3.1        & 4.3          & 33         & 2,  0.56     & 19, 0.205         &  4.0   & 2.5    & 0.82   & 2.6       &  5.2      & 9  \\
France & 28/02/2020 -- 03/05/2020 & 1.3        & 2.9        & 5.0          & 29         & 2,  0.60     & 17, 0.195         &  3.8   & 2.28   & 0.74   & 2.8       &  5.3      & 12 \\
UK     & 01/03/2020 -- 10/05/2020 & 1.28       & 2.8        & 6.2          & N/A        & 15, 0.65     & 25, 0.270         &  3.6   & 2.34   & 0.97   & 3.4       &  5.8      & 10
\end{tabular}
\end{center}
\end{table*}

\begin{figure}[tb]
\centering
\includegraphics[width=\columnwidth]{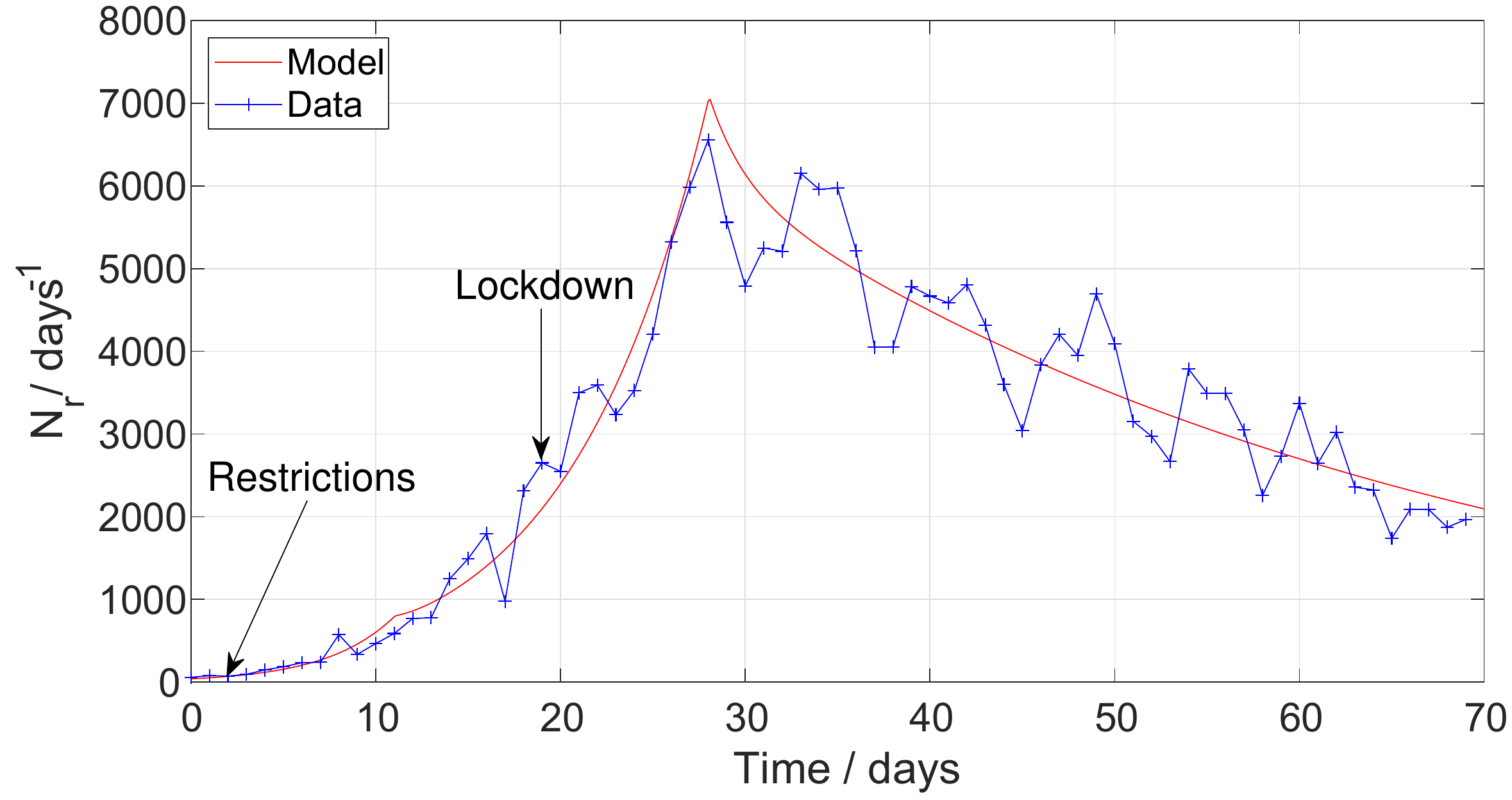}
\caption{Italian outbreak validation: new daily reported cases}
\label{fig:ItalyNr}
\end{figure}\begin{figure}[tb]
\centering
\includegraphics[width=\columnwidth]{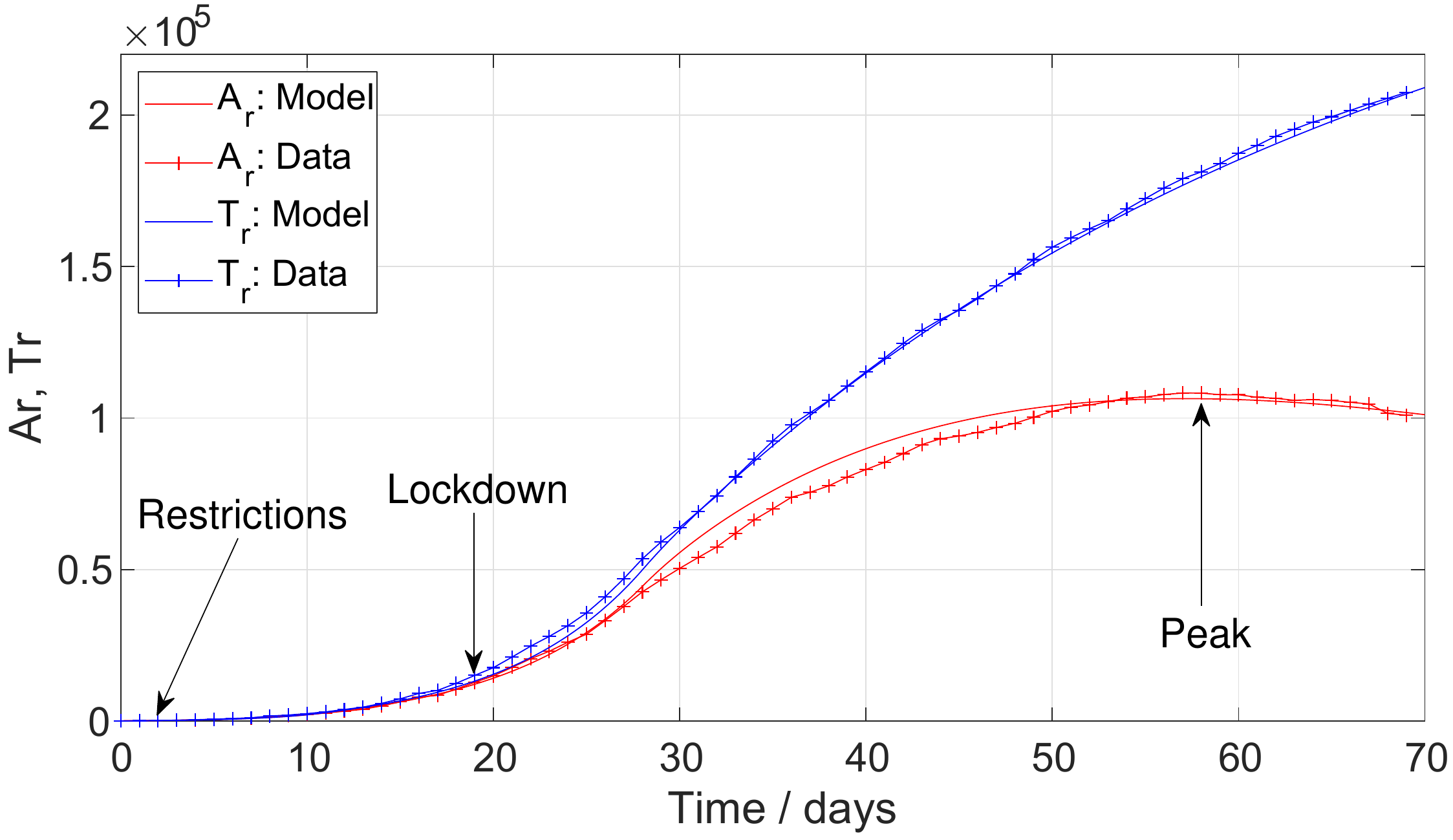}
\caption{Italian outbreak validation: active and total reported cases}
\label{fig:ItalyArTr}
\end{figure}

The detailed results for the case of Italy, computed with the code available online in \cite{Covid19Control}, are reported in Figs. \ref{fig:ItalyNr}-\ref{fig:ItalyArTr}. The interplay between $E_t$ and $R_t$ accounts for the growth or decay of cases, depending on $\R_t$. The $E$ compartment is essential to explain why $N_r(t)$ does not drop sharply, when $\beta$ is sharply reduced, once the $\tau_m$ delay has elapsed, while the $L$ compartment is necessary to explain the much slower dynamic response of active reported cases $A_r(t)$. The validation results of the other cases are reported in \cite{Casella2020}.

\section{Control}
\label{Sec:Control}
The effects of the application of the two control policies outlined in the Introduction will now be analysed. The title of this section may well be "Respect the Unstable" \cite{Stein2003}: feedback control strategies should not be applied light-heartedly to safety-critical unstable systems.

\subsection{Suppression}
\label{Sec:Suppression}
The suppression strategy can be brutally summarized in the following terms: as soon as $A_r(t)$ reaches a value $A_s$ which is scary enough to decision makers to overcome their reluctance to disrupt the social and economic life of their country, drastic containment measures are taken:
\begin{equation}
u(t) = \left\{ \quad \begin{matrix}
0, &   A_r(t) < A_s \\
\bar{u}, &  A_r(t) \geq A_s
\end{matrix} \right. .
\end{equation}

If the threshold $A_s$ is crossed at time $t_s$ and $\bar{u}$ is large enough, then $\beta(\bar{u}) / \gamma < 1$ and thus $r < 0$. Assuming also $d(t)$ remains constant for $t \geq t_s$, Eqs. \eqref{eq:Etf}--\eqref{eq:Ltf} form a homogeneous LTI system with three negative eigenvalues $r$, $p$, and $-\delta$. The actual number of eventually tested positive exposed subject $E_t(t)$ will start decaying immediately; however, the number of new daily reported infectious cases $N_r(t)$ will continue its exponential growth for $\tau_m$ days, before starting to decay as well. The number of reported active cases $A_r(t)$, hence the required number of beds in hospitals and intensive care units, will also stop increasing exponentially after $\tau_m$ days, but will continue growing and peak much later, due to the much slower time constant $1/\delta$, see, e.g., Fig. \ref{fig:ItalyArTr}. Then, states and outputs asymptotically approach the equilibrium in the origin, that corresponds to the eradication of the virus.

The peak value of active cases $A_p = \max A_r(t)$ can be computed by numerical integration of Eqs. \eqref{eq:Etf}--\eqref{eq:Ltf} and \eqref{eq:Arf}. The ratio $M = A_p/A_r(t_s)$ can be quite large, e.g. $M = 8$ for the Italian outbreak and $M = 10$ for the French outbreak.

Assuming that a fraction $\sigma$ (about $4\%$ in Italy) of active cases requires intensive care, and that $N_{ic}$ intensive care beds are available, a wise choice of $A_s$ requires $\sigma A_r(t) < N_{ic} \enspace \forall t$; hence, $A_s < N_{ic}/\sigma M$. Political decision-makers without a training in mathematical modelling may have difficulties in understanding the role and magnitude of factor $M$ and may be caught by surprise once it is too late to act.

\subsection{Mitigation}
\subsubsection{Policy statement}

The basic idea of mitigation policies is to manage the outbreak, in particular the trajectory of $A_r(t)$, in order to avoid overloading the public health system, without trying to suppress it. This strategy was followed until 16 March 2020 by the UK government, which aimed at achieving herd immunity \cite{Hunter2020}, and until at least 10 June 2020 by the Swedish government \cite{Paterlini2020}.

\subsubsection{Mathematical formalization}

The first step to enact this strategy is to compute a reference  control policy $u^0(t)$, obtained by the application of suitable NPIs over time, whose effects on $\beta$ is accurately calibrated, leading to reference trajectories $N^0_r(t)$, $A^0_r(t)$ and $T^0_r(t)$ for the corresponding indicators, respecting the constraint  $\sigma A_r^0(t) < N_{ic}$. These trajectories can be obtained by means of constrained optimal control, using sophisticated models of the epidemic as, e.g., the ones reported in \cite{Ferguson2020a}, \cite{GattoEtAl2020}, and \cite{GiordanoEtAl2020}.

\begin{figure}[tb]
\centering
\includegraphics[width=\columnwidth]{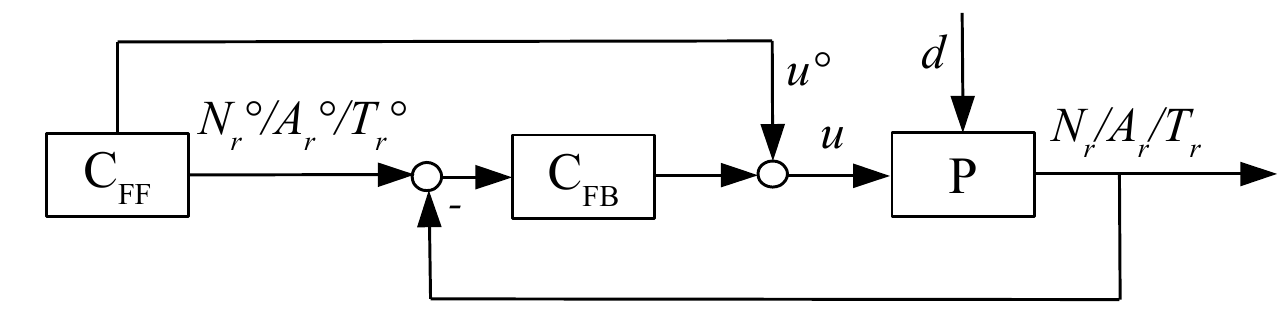}
\caption{Mitigation control architecture}
\label{fig:control_diagram}
\end{figure}

The unstable nature of the state trajectories while $r > 0$ makes an open-loop implementation of this policy infeasible, unless one wants to risk runaways that can cause the collapse of the public health system. The reference trajectory should rather be followed by adapting the public policy measures $u(t)$ in real time, based on the values of $N^r(t)$, $A^r(t)$, and $T^r(t)$, which are constantly monitored by the authorities. This corresponds in principle to closing a \emph{feedback loop} to stabilize the unstable reference trajectory, see Fig. \ref{fig:control_diagram}. 

\subsubsection{Ideal Feedback controller design}
The process model \eqref{eq:Etf}-\eqref{eq:Arf} can be linearized around the reference trajectory at time $t_a$, obtaining a linear model with constant coefficients except for the terms $\beta(u^0(t_a),d(t_a))$ and $\frac{\partial \beta(u^0(t_a),d(t_a))}{\partial u}$, which depend on $t_a$ for non-trivial reference control trajectories $u^0(t)$. For the sake of the subsequent analysis, we assume that these parameters change over a time scale which is much longer than the time scale of the closed-loop system feedback response, a common assumption when dealing with gain-scheduling control, and thus consider them as constants, with the value they have at time $t_a$ around which the feedback stability analysis is performed. The transfer functions of the linearized process then reads:
\begin{align}
\frac{\Delta N_r(s)}{\Delta u(t)} &= \frac{\mu(t_a)(s+\gamma)}{(s-p(t_a))(s-r(t_a))} e^{-\tau_m s} \label{eq:Nr_s} \\
\frac{\Delta A_r(s)}{\Delta u(t)} &=  \frac{\mu(t_a)(s+(\gamma + \delta))}{(s-p(t_a))(s-r(t_a))(s+\delta)} e^{-\tau_m s} \label{eq:Ar_s} \\
\frac{\Delta T_r(s)}{\Delta u(t)} &= \frac{\mu(t_a) (s+\gamma)}{(s-p(t_a))(s-r(t_a))s} e^{-\tau_m s} \label{eq:Tr_s} 
\end{align}
where $\mu(t_a) = \epsilon \frac{\partial \beta(u^0(t_a),d(t_a))}{\partial u}I^0(t_a)$, and $p(t_a)$ and $r(t_a)$ are the eigenvalues of system \eqref{eq:Etf}-\eqref{eq:Itf} linearized at $t = t_a$ around the reference trajectory.

By making the very optimistic assumptions that the parameters $\epsilon$, $\gamma$, $\delta$, and $\tau_m$ are constant and perfectly known, that the function $\beta(u,d)$ that expresses the effects of public policy decisions is perfectly known, monotonously decreasing and smooth with respect to $u$, and that $d(t)\equiv 0$, one can design the feedback controller $\mathrm{C_{FB}}$ as a linear controller $C(s)$ with gain scheduling, that compensates for the nonlinearity of the process gain, resulting in a linear and (approximately) time-invariant loop dynamics. While doing so, one should also account for an additional delay $\tau_c$ of $2 \div 4$ days within the controller, corresponding to the decision making and implementation delay. Assuming one wants to use $N_r(t)$ for feedback control, the overall control law is thus:
\begin{align}
\label{eq:PD}
& u(t) = u^0(t) + \frac{1}{\mu (t-\tau_c)} u_f(t-\tau_c) \\
& u_f(s) = C(s) \left[N_r^0(s)-N_r(s) \right]
\end{align}
and the loop transfer function of the controlled system is
\begin{equation}
L(s) = C(s) \frac{\mu_r(s+\gamma)}{(s-p(t-\tau_c))(s-r(t-\tau_c))} e^{-s(\tau_m + \tau_c)},
\end{equation}
where $\mu_r$ is the ratio between the \emph{actual} value of the gain $\mu$ of transfer function \eqref{eq:Nr_s} and its \emph{reference} value used for gain scheduling. In ideal conditions, $\mu_r = 1$, though results from \cite{Flaxman2020} imply this gain is subject to significant uncertainty. In case one wants to control $A_r(t)$ or $T_r(t)$, similar consideration apply, using \eqref{eq:Ar_s} or \eqref{eq:Tr_s} instead of \eqref{eq:Nr_s}.

In all the three cases, the loop transfer function reveals the very dangerous nature of this process, which features a time delay $\tau$, an uncertain gain $\mu_r$, and an unstable pole with time constant $T$ if $\beta(t_a)/\gamma > 1$,  where
\begin{align}
&T = \frac{1}{r} = \frac{T_d}{log(2)} \\
&\tau = \tau_t + \tau_r + \tau_c.
\end{align}

Considering the values reported in Table \ref{tab:parameters}, $T_d = 4.3 \div 5.8$ before lockdown, while $\tau = 9 \div 12$ days.

\subsubsection{Control feasibility}
\label{sec:control_feasibility}

In order to guarantee some robustness of the system performance against the large gain uncertainty of the process, the Bode plot of $\abs{L(j\omega)}$ should maintain a roughly constant slope over a sufficiently wide interval around the crossover frequency $\omega_c$, thus approximating Bode's ideal loop transfer function.

When $\R_t(t) > 1$, hence $r(t)>0$, the analysis reported in \cite{Astrom2000}, Sect. 4.6, leads to conclude that if one wants to limit the maximum norm of the sensitivity function $M_s < 2$ to achieve some robustness, the product of the unstable pole $r$ and of the time delay $\tau$ should be $r\tau < 0.156$. Introducing the doubling time $T_d$, this condition becomes:
\begin{equation}
\label{eq:controllability1}
\frac{\tau}{T_d} < 0.225,
\end{equation}
i.e., under very optimistic assumptions on the knowledge of the process parameters, feedback control is feasible only if the overall loop delay is less than a quarter of the doubling time of the outbreak. Considering the values reported in Table \ref{tab:parameters}, this limitation translates into $\R_t(t) < 1.1$.

When $\R_t(t) < 1$, hence $r(t)<0$, \cite{Astrom2000} concludes that the maximum crossover frequency is $1.57/\tau$; hence, the time constant of the response of the feedback controller to unexpected disturbances cannot be less than $T = \tau / 1.57$.

Unfortunately there is no theorem that can be directly invoked to prove that \emph{any} feedback control policy would not suffer from the same limitations of a carefully scheduled linear controller. However, the results of this analysis provides an insightful benchmark, pointing out the crucial role of measurement and decision delays, which should explicitly be taken into account for feedback control design, and minimized as much as possible. It also suggest that robust feedback control may not be feasible around trajectories where $\R_t$ is significantly above one. For example, this may explain the runaway scenarios happening with non-negligible likelihood in \cite{BinEtAl2020}, when taking into account the statistical distribution of the uncertain process parameters. 

\section{Conclusions}
\label{sec:Conclusions}
Governments all the world over are faced with very challenging life-or-death decisions regarding the management of the COVID-19 epidemic, involving the balance between public health and economic issues. In order to take such decisions, they rely on expert advice based on the results of epidemiological mathematical models and on daily case reports, based on swab test results.

This paper puts the problem in a control systems perspective as a feedback control problem, using a simple model to capture the control-relevant dynamic response of those reports to the application of NPIs. The model was tuned and validated with data from four different outbreaks.

These are the main results of the analysis:
\begin{itemize}
\item The suppression strategy is effective if NPIs are strong enough to obtain $R_t < 1$, but it requires to understand the role of the multiplicative factor $M$ to correctly decide when it is the right time to enforce them.
\item Mitigation strategies are limited by the combination of delay, uncertainty, and unstable dynamics. Designing robust stabilizing controllers around trajectories with $\R_t > 1.1$ is likely to be difficult or impossible. Reducing the overall delay by 50\% would bring the limit to $\R_t > 1.2$. This information is particularly relevant for the management of the reopening phase after lockdown.
\item Measurement and decision delays play a crucial role in determining the feedback control performance and stability; hence, they should be explicitly taken into account in the design of \emph{any} feedback controller, and minimized as much as possible, e.g., by promoting fast testing policies and technologies.
\item The analysis and design of NPIs can benefit from control theory tools, possibly suggesting viable solutions or pointing out shortcomings of proposed strategies, that are not obvious to epidemiologists and physicians.
\end{itemize}

At the time of this writing, the emerging consensus seems to be that the safest policy to address exponentially growing COVID-19 outbreaks is to apply aggressive enough suppression policies; the results reported in this paper can further motivate why this is actually the case.

These results could also be useful to devise effective and safe strategies to cope with the reopening phase that countries face after successfully suppressing the first outbreak, in particular if the number of new daily cases becomes too large to allow for testing, tracing and tracking of individual cases.


\bibliographystyle{IEEEtranDOI}
\bibliography{paper}

%

\section*{Supplementary material}
Detailed results of the model validation in the cases of China, France, and UK are discussed in this appendix, which does not fit the strict 6-page limit of IEEE Control Systems Letters publications, and therefore is only made available in the arXiv version of the paper.

\subsection{China}

\begin{figure}[b]
\centering
\includegraphics[width=\columnwidth]{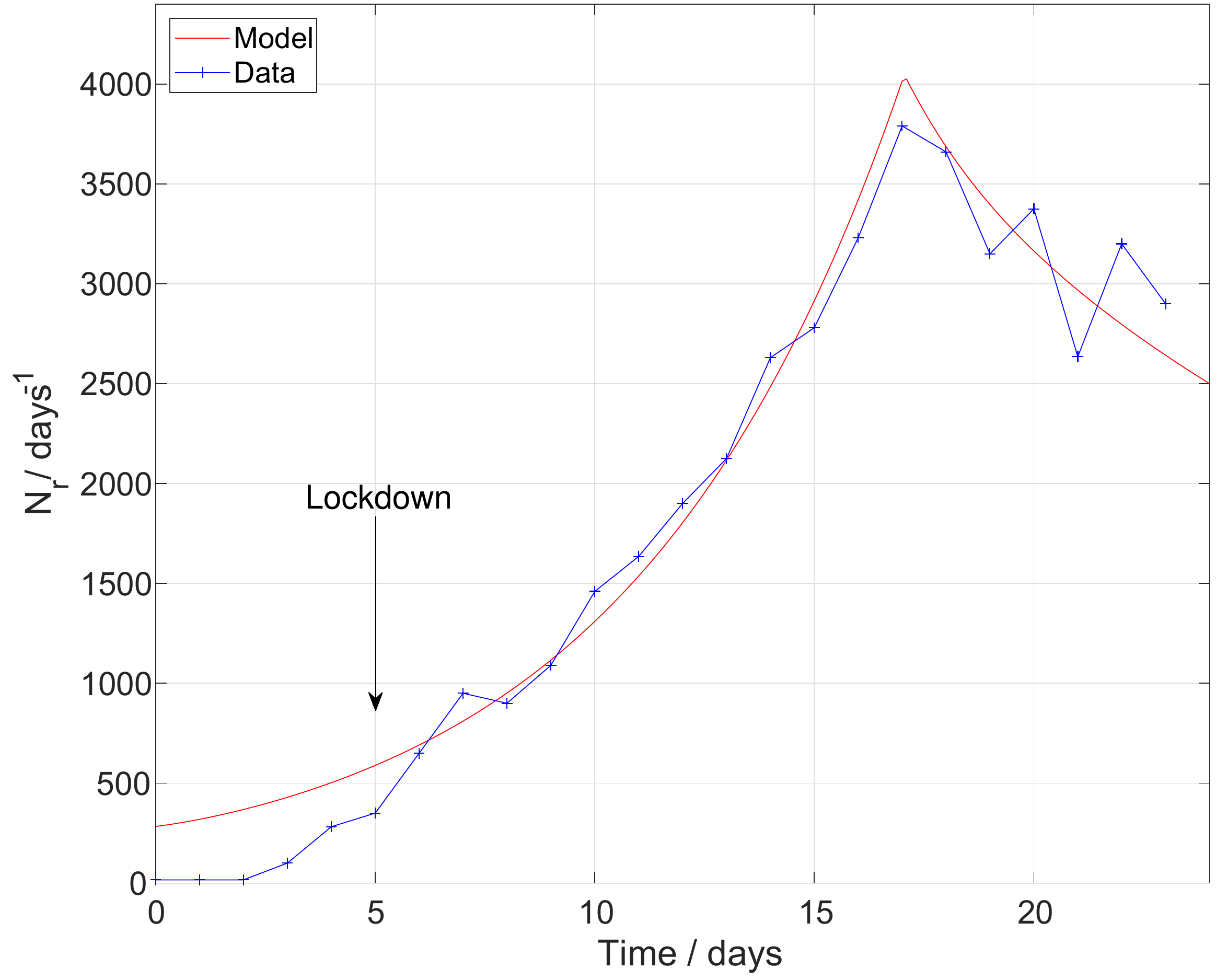}
\caption{Chinese outbreak validation: new daily reported cases}
\label{fig:ChinaNr}
\end{figure}

Fig. \ref{fig:ChinaNr} reports the validation of the $N_r(t)$ trajectory for the Chinese outbreak, based on data from \cite{JAMA}. 

The lockdown became effective on day 5 for the city of Wuhan, which had the majority of cases at the time, and on day 6 for all the Hubei region. Since it is not possible from available data to figure out what are the separate effects of two such very close events, the second one was lumped together with the first one, which produced the largest number of reported cases.

The data at the beginning of the time series until day 5 doesn't really fit well an exponential curve, but that is probably due to issues with the collection of swab test results. However, the effect of the time delay $\tau_m$ is very clearly visible, as the number of new reported daily cases only stopped growing exponentially after 12 days. The behaviour of $N_r(t)$ after the peak allowed to tune the $\gamma$ and $\epsilon$ parameters, and to estimate a reproduction number $\R_t = 0.64$ immediately after the effect of lockdown was felt.

\subsection{France}
\begin{figure}[tb]
\centering
\includegraphics[width=\columnwidth]{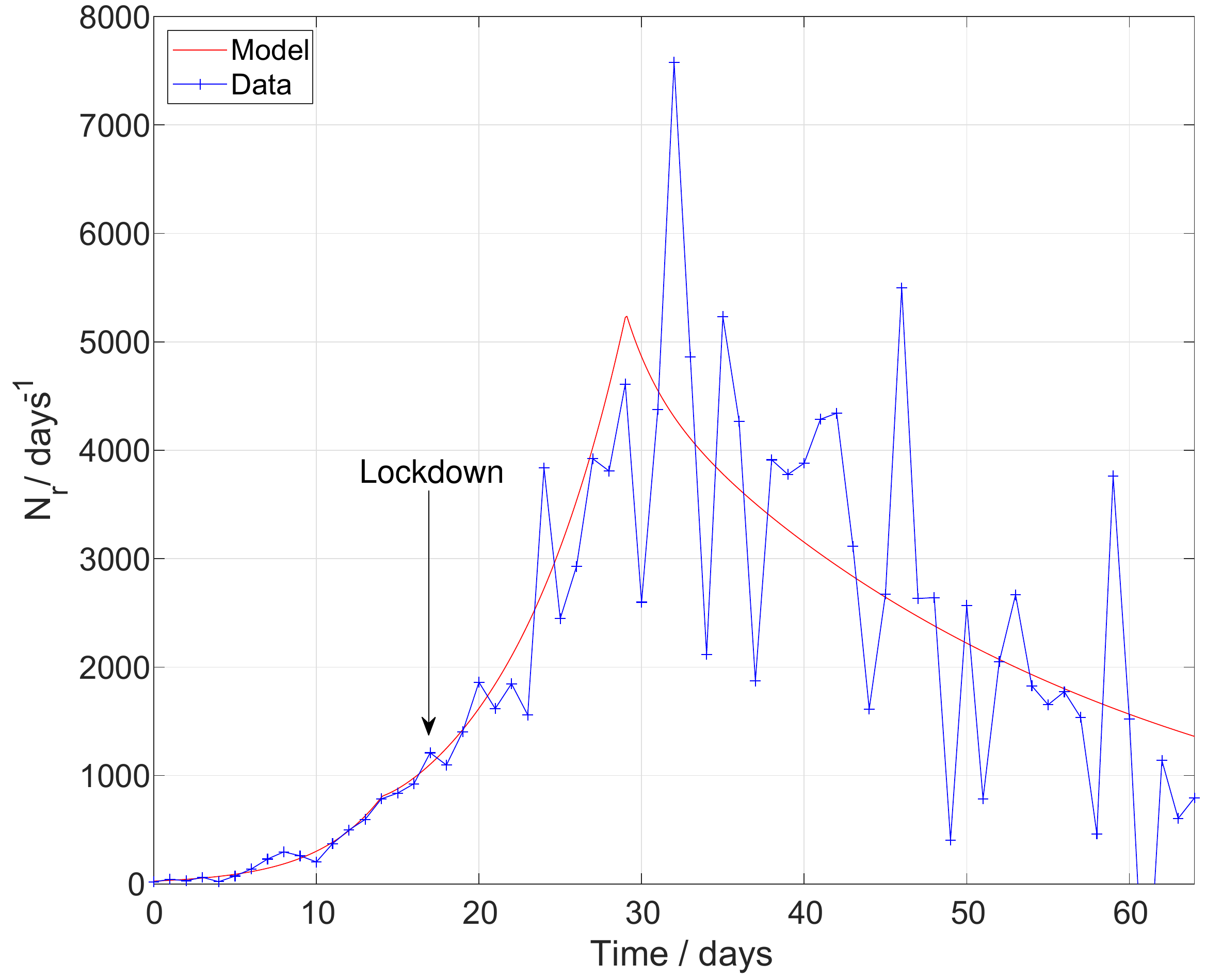}
\caption{French outbreak validation: new daily reported cases}
\label{fig:FranceNr}
\end{figure}

Fig. \ref{fig:FranceNr} reports the validation of the $N_r(t)$ trajectory for the French outbreak, based on data from \cite{Worldometer}.

\begin{figure}[tb]
\centering
\includegraphics[width=\columnwidth]{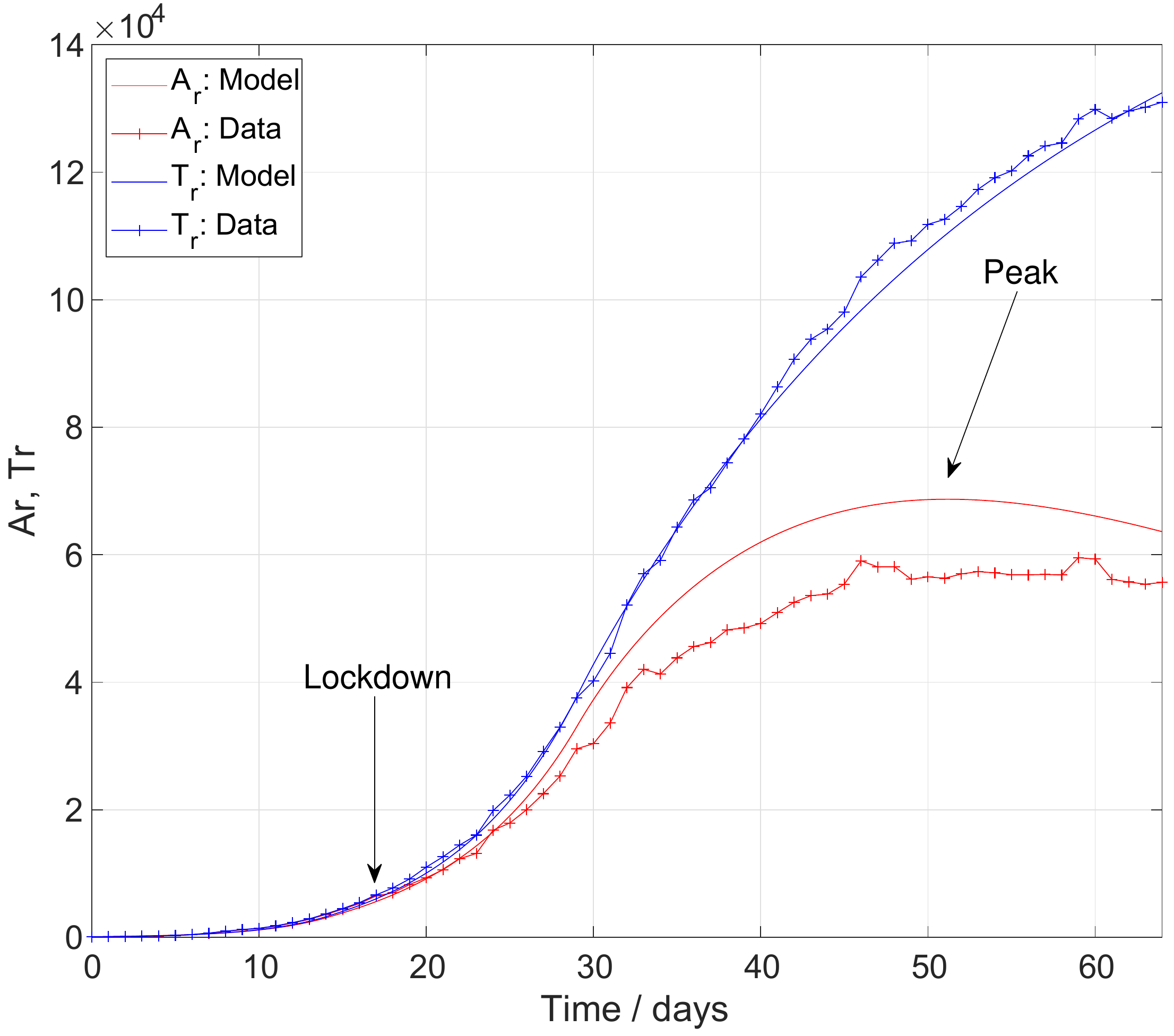}
\caption{French outbreak validation: active and total reported cases}
\label{fig:FranceArTr}
\end{figure}

As already mentioned in Sect. \ref{sec:Validation}, the doubling time increased significantly after day 14, corresponding to a change of $\beta$ happened around day 2, given the delay $\tau = 12$. The cause of this change is unknown, it may actually have been caused by some feedback effects on social behaviour caused by the growing fear of contagion sparked by the nearby Italian case. The trend of $N_r(t)$ after the delayed effect of the lockdown on day 17 is very noisy but clearly decreasing on average. This is confermed by looking at the very good match in the validation of the $T_r(t)$ trajectory reported in Fig. \ref{fig:FranceArTr}: as $T_r(t)$ is the integral of $N_r(t)$ (see \eqref{eq:Trf}), high-frequency noise is filtered out.

As to the trajectory of active cases $A_r(t)$ shown in Fig. \ref{fig:FranceArTr}, the matching of the simulation results with the data is very good for the first 25 days, then some mechanism seems to have somewhat dampened the experimental curve more efficiently than the model in this paper. In any case, the model overestimates the peak value by around 15\%, which is not that bad, considering how simple the model is. 

\subsection{UK}
Finally, we take into consideration the case of UK, again using data taken from \cite{Worldometer}. The virus was basically allowed to run unchecked until March 16, 2020 (day 15), when the government issued some recommendations to avoid going to pubs or theaters, without however closing them or issuing any legally binding rules or restriction. This seems to have slowed down the exponential growth after a 10-day delay, increasing the doubling time from 3.4 to 5.8 days. 

\begin{figure}[tb]
\centering
\includegraphics[width=\columnwidth]{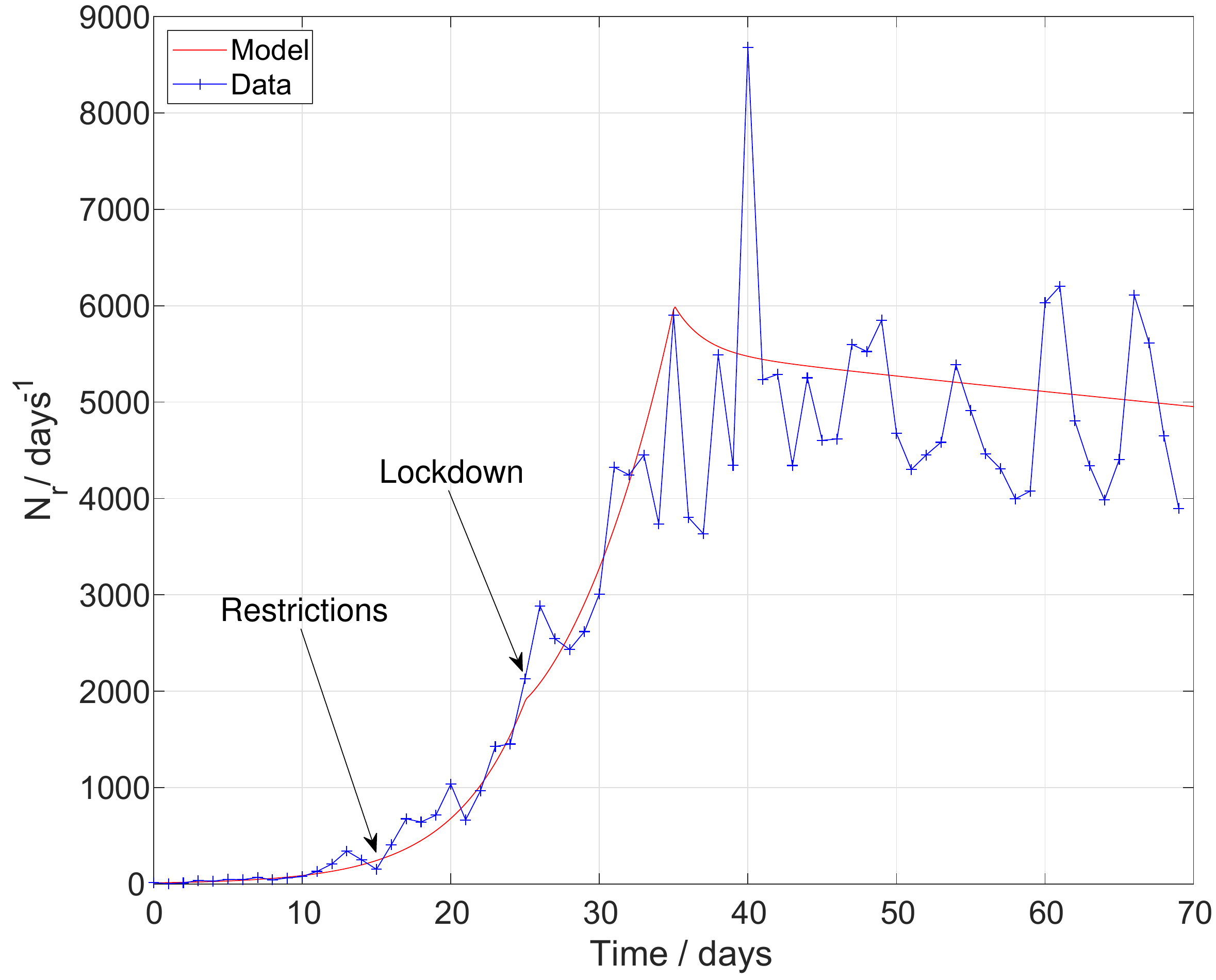}
\caption{UK outbreak validation: new daily reported cases}
\label{fig:UKNr}
\end{figure}

Full lockdown was announced on Mar 23 and went into effect on Mar 26 (day 25). Also in this case, the effect on $N_r(t)$ was felt after a significant delay, in this case of about 10 days, see Fig. \ref{fig:UKNr}. Once that time interval was elapsed, the number of new daily positive swab tests plateaued, instead of starting to decay, settling down on a trajectory with estimated $\R_t = 0.97$, very close to the stability limit. 

\begin{figure}[tb]
\centering
\includegraphics[width=0.92\columnwidth]{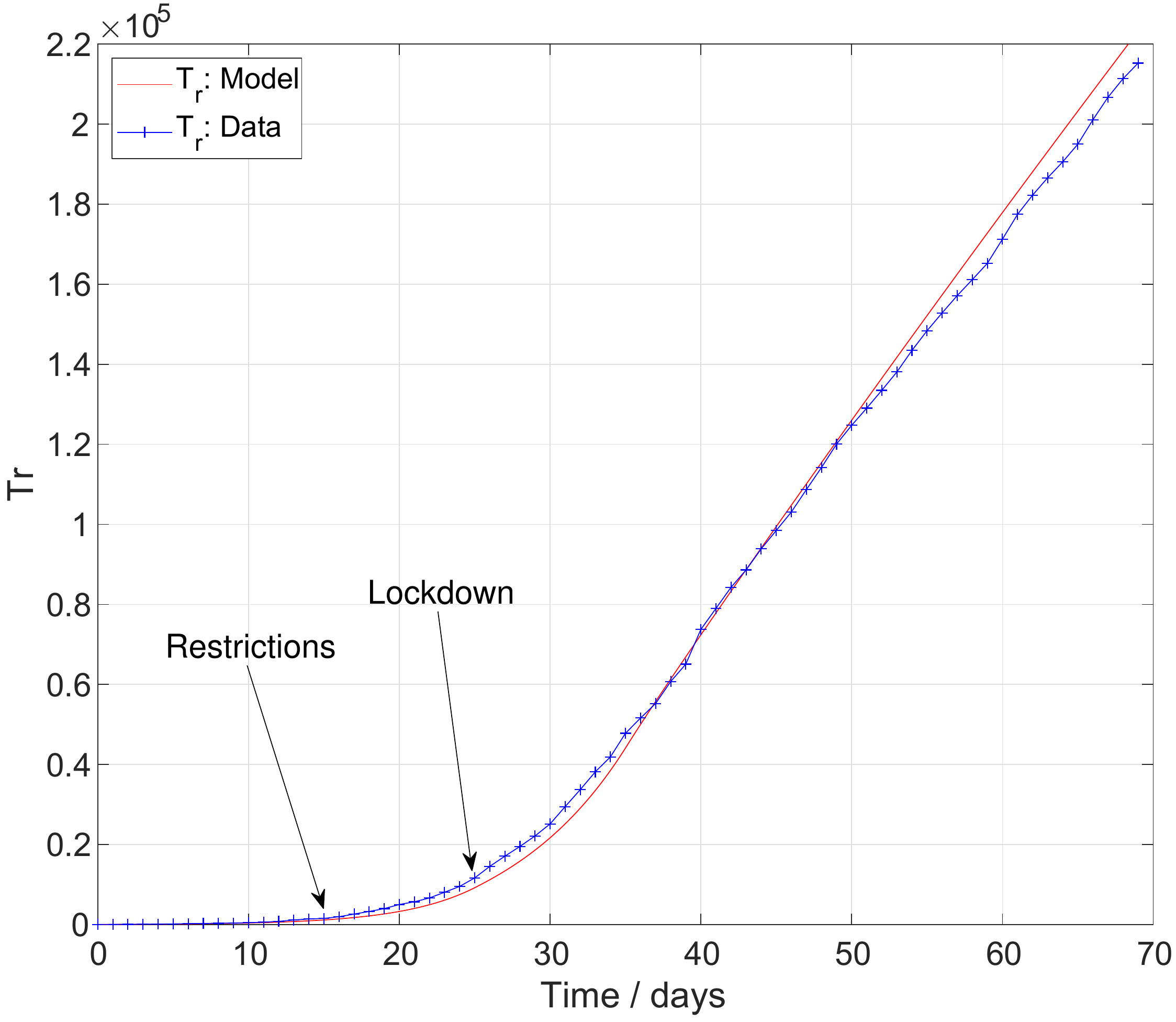}
\caption{UK outbreak validation: active and total reported cases}
\label{fig:UKTr}
\end{figure}

Also in this case, the experimental data about $N_r(t)$ are very noisy; however, an excellent match with data is obtained with the cumulated $T_r(t)$ trajectory, which is less sensitve to nois because of the additional integral effect, see Fig. \ref{fig:UKTr}. Unfortunately, data for $A_r(t)$ were not available from \cite{Worldometer}, so it was not possible to validate the active cases output in this case.

\newpage

\enlargethispage{-15in}

\end{document}